\documentstyle[preprint,aps]{revtex}
\makeatletter
\newcommand{\bc}{\begin{center}}
\newcommand{\ec}{\end{center}}
\newcommand{\be}{\begin{eqnarray}}
\newcommand{\ee}{\end{eqnarray}}
\newcommand{\nn}{\nonumber}

\def\bsg{b\to s \gamma}

\def\bsgl{b\to s g}

\def\figcap{\section*{Figure Captions\markboth
     {FIGURECAPTIONS}{FIGURECAPTIONS}}\list
     {Fig. \arabic{enumi}:\hfill}{\settowidth\labelwidth{Fig. 999:}
     \leftmargin\labelwidth
     \advance\leftmargin\labelsep\usecounter{enumi}}}
 \relax
\def\reflist{\section*{References\markboth
     {REFLIST}{REFLIST}}\list
     {[\arabic{enumi}]\hfill}{\settowidth\labelwidth{[999]}
     \leftmargin\labelwidth
     \advance\leftmargin\labelsep\usecounter{enumi}}}
 \relax

\begin{document}
\draft
\thispagestyle{empty}
\centerline{hep-ph/yymmnn}\vskip -.3cm
\centerline{\hfill  NHCU-HEP-94-17}
\centerline{\hfill  NTUTH-94-16}
\vskip 0.5cm
\begin{center}
\Large
Limit on $Br(b\to s g)$ in Two Higgs Doublet Models
\end{center}
\vskip 1cm

\centerline{
Chao-Qiang Geng and Paul Turcotte
}\vskip 0.3cm
\centerline{
Physics Department,
National Tsing Hua University, Hsinchu, Taiwan 30043, R.O.C.}
%
\vskip 0.9cm
\centerline{
Wei-Shu Hou
}\vskip 0.3cm
\centerline{
Physics Department,
National Taiwan University, Taipei, Taiwan 10764, R.O.C.\footnotemark
}
\vskip -0.1cm
\centerline{and}
\vskip -0.1cm
\centerline{
Theory Division, CERN, CH-1211,
Geneva 23, Switzerland}

\vskip 1.0cm
\begin{center}
 {\large\bf Abstract}
\end{center}
\begin{quote}

Using the recent CLEO measurement of $Br(b\to s \gamma)$,
we find that the branching ratio of $b\to s g$
cannot be larger than $10\%$ in two Higgs doublet models.
The small experimental value of $Br(b\to e\bar{\nu}X)$
can no longer be explained by charged Higgs boson effects.
\end{quote}
\vfill
\footnotetext{
          Permanent address.}
%

\newpage
\narrowtext

It is well known that the process $b\to s\gamma$ is extremely sensitive
to new physics beyond the standard model, in particular, that containing
a charged Higgs boson. In 1993, the CLEO collaboration placed
an upper limit of $Br(b\to s\gamma) < 5.4\times 10^{-4}$
on the inclusive branching ratio \cite{CLEO},
which has inspired a large number of studies of this decay
in various models for new physics \cite{Hewett-r}.
Stringent constraints are obtained.
Recently, CLEO has {\it measured} the inclusive branching ratio
to be \cite{CLEO2} $(2.32\pm 0.51 \pm 0.29 \pm 0.32)\times 10^{-4}$,
in short, $(2.3 \pm 0.7) \times 10^{-4}$, corresponding to the
$95\%$ confidence level range of
\begin{equation}
1 \times 10^{-4} < Br(b\to s\gamma)< 4\times 10^{-4}.
\end{equation}

In the standard model, there is a surprisingly large
QCD enhancement of $b\to s\gamma$ \cite{qcdi} amplitude.
This has stimulated intense efforts in
calculating QCD corrections to leading order (LO),
as well as partial calculations to next-to-leading order (NLO)
\cite{Buras,Ciuchini}.
The CLEO result of eq. (1) is not far from the prediction of
$Br(b\to s\gamma)$ in the standard model.
This implies that not much room is left for new physics contributions
to other $b\to s$ transitions such as $b\to s g$, where the
emitted gluon is ``on-shell".

It was shown in ref.  \cite{HW} that
for some choices of parameters in
two Higgs doublet models, charged Higgs boson effects
may enhance the decay branching ratio of $b\to s g$
beyond the $10\%$ level.
Grz\c{a}dkowski and Hou \cite{GH} have pointed out
that if $\bsgl$ rate is at the $(10-20)\%$ level,
the discrepancy on $Br(b\to e \nu X)$
between experimental measurement  $(10.7\pm.5\%)$ \cite{benx-expt}
and theoretical expectations $(>12\%$ in the standard model) \cite{benx-th}
could be resolved.
It is therefore of interest to check whether
the possibility of $Br(b\to s g)\sim 10\%$ still holds
once one includes the constraint imposed by eq. (1).

In this report we focus on two Higgs doublet models (2HDM).
These models are very simple extensions of the standard model,
yet they exhibit some of the characteristics of a more
complicated scalar structure typical of most theories
beyond the standard model.
We will consider the two distinct models (I and II)
that naturally avoid tree-level FCNCs \cite{GW}.
In Model I, one doublet $(\phi_2)$ couples to all fermions
and the other $(\phi_1)$ decouples from the fermion sector.
In Model II, $\phi_2$ couples to up-type quarks while
$\phi_1$ couples to down-type quarks.
This type of model occurs in minimal realization of supersymmetry,
or in models with a Peccei-Quinn symmetry \cite{hunter}.
The major non-standard feature of these models
is the appearance of extra physical scalar fields.
We consider only the effect of charged Higgs bosons.
Two parameters are sufficient to account for the additional effects.
We take these to be $m_H$, the mass of the charged Higgs boson,
and $\xi \equiv v_1 / v_2$, the ratio of the vacuum
expectation values of $\phi_1$ and $\phi_2$.
Note that $\xi=1/\tan\beta$,
as is commonly used in supersymmetric models.
In 2HDM,
the charged Higgs couple to quarks with the same
quark mixing matrix as the standard charged current.

The standard model calculation for $\bsg$,
including up to date QCD corrections,
can be found, for example, in Refs. \cite{Buras,Ciuchini}.
The inclusive branching ratio is given by
\begin{eqnarray}
Br(b\to s \gamma)& =& {\Gamma(b\to s\gamma)\over
\Gamma(b\to c e\bar{\nu})}\; Br(b\to c e\bar{\nu})
\nonumber \\
&=& {\vert V_{ts}^*V_{tb}\vert^2 \over \vert V_{cb}\vert^2}
{6\alpha\over \pi f(m_c/m_b)}\;{
1\over \Omega (m_t/M_W,\mu)}
\;\vert C_7^{eff}(\mu)\vert^2
\;Br(b\to c e\bar{\nu}),
\label{eq:bsA}
\end{eqnarray}
where the Wilson coefficient
\be
C^{eff}_7(\mu) &=& \eta^{{16\over 23}} C_7(M_W) +
{8\over 3} \left( \eta^{{14\over 23}} - \eta^{{16\over 23}}
\right) C_8(M_W)  + C_2(M_W) \sum_{i=1}^8 a_i \eta^{b_i},
\ee
includes short distance effects at $M_W$ scale,
while perturbative QCD effects are accumulated when running
down to the physical scale $\mu$,
with
$\eta = \alpha_S(M_W)/\alpha_S(\mu)$.
In eq. (\ref{eq:bsA}), the phase space factor $f(z)$ is given by
\be
f(z) &=& 1 - 8z^2 + 8z^6 - z^8 - 24z^4 \ln z,
\ee
and the quantity $\Omega (z)$ contains
the $O(\alpha_S)$ QCD corrections to the
semileptonic decay rate \cite{c24,c25}
and is given by 
\begin{equation}
\Omega (x, \mu)  \simeq  1-{2\alpha_S(\mu)\over 3\pi}
\left[\left( \pi^2 -{31\over 4} \right) (1-x)^2 + {3\over 2} \right].
\label{eq:omega}
\end{equation}
The scheme-independent numbers $a_i$ and $b_i$ are given by
\cite{Ciu}
\be
a_i &=& \matrix{
({626126\over 272277}, &-{56281\over 51730}, &-{3\over 7}, & -{1\over 14},
        &     -0.6494,       &      -0.0380,     & -0.0186,  &  -0.0057) },
\nn\\
b_i& =& \matrix{ ({14\over 23}, & {16\over 23}, & {6\over 23},&-{12\over 23},
        &      0.4086,       &      -0.4230,     & -0.8994,  &   0.1456 ) },
\ee
respectively.
Defining $x=m_t^2/M_W^2$, $h=m_t^2/M_H^2$,
the coefficients $C_i(M_W)$ are \cite{HW}
\be
C_2(M_W)^I & =& 1,
\nn\\
C_7(M_W)^I &=& -{1\over 2} A(x)+
\xi^2
\left[B(h)-{1\over 6}A(h)\right],
\nn\\
C_8(M_W)^I &=& -{1\over 2} D(x)+
\xi^2
\left[E(h)-{1\over 6}D(h)\right],
\ee
for Model I, and \cite{HW,wise}
\be
C_2(M_W)^{II} & =& 1,
\nn\\
C_7(M_W)^{II} &=& -{1\over 2} A(x)
-B(h)-{1\over 6}\xi^2A(h),
\nn\\
C_8(M_W)^{II} &=& -{1\over 2} D(x)
-E(h)-{1\over 6}\xi^2D(h),
\ee
for Model II, where
\be
A(x) &=& {-x\over 12(1-x)^4}
\left[6x(3x-2)\ln x+(1-x)(8x^2+5x-7)\right],
\nn\\
D(x) &=& {x\over 4(1-x)^4}\left[6x\ln x-(1-x)(x^2-5x-2)\right],
\nn\\
B(x) &=& {x\over 12(1-x)^3}
\left[(6x-4)\ln x+(1-x)(5x-3)\right],
\nn\\
E(x) &=& {-x\over 4(1-x)^3}\left[2\ln x+(1-x)((3-x)\right].
\ee

Analogously,
the branching ratio for $b\to s g$, where the gluon is
{\it on-shell} (in the sense of a ``gluon jet"),
can be written as,
\be
Br(b\to s g)
&=& {\vert V_{ts}^*V_{tb}\vert^2 \over \vert V_{cb}\vert^2}
 {8\alpha_S(\mu)\over \pi f(m_c/m_b)}\;{1\over \Omega(m_t/M_W,\mu)}
\;\vert C_8^{eff}(\mu)\vert^2
\;Br(b\to c e\bar{\nu}),
\label{eq:bsg}
\ee
where \cite{Buras}
\be
C^{eff}_8(\mu) &=& \left[C_8(M_W)  + {313063\over 363036}\right]
\eta^{{14\over 23}}-0.9135 \eta^{0.4086}
\nn\\
&&   +  0.0873 \eta^{-0.4230}
              -0.0571 \eta^{-0.8994}  -  0.0209 \eta^{0.1456}\:.
\label{eq:c8}
\ee
Notice the explicit $\mu$-dependence of eq. (\ref{eq:bsg})
on $\alpha_S(\mu)$. The $\mu$ scale of this $\alpha_S$
does not have to be the same as that of $C^{eff}_8(\mu)$,
but we treat them as if they are the same. From
eqs. (\ref{eq:bsA}) and (\ref{eq:bsg}), we form the ratio
\be
R\ & \equiv\ & {Br(\bsgl)\over Br(\bsg)} \:=\:
{4\over 3}\;{\alpha_S(\mu)\over \alpha}
\;\left\vert{C_8^{eff}(\mu) \over C_7^{eff}(\mu) }\right\vert^2,
\label{eq:R}
\ee
which is independent of $m_c/m_b$.

The scale $\mu$ denotes the renormalization scale of
the effective $\bsg$ Hamiltonian.
It should be of order $m_b$, but need not be exactly
equal to $m_b$, and we shall take it to be in the range
between 2.5 to 10 GeV as used in Ref. \cite{Buras,Ali-mu}.
For simplicity we have taken it to be the same
as the scale at which the parameter $\alpha_S$ is expanded
for the QCD corrections to the semileptonic decay rate,
eq. (\ref{eq:omega}).
We express $\alpha_S(\mu)$
in terms of its value at $\mu = M_Z$, i.e.
\be
{\alpha_S(M_Z)\over \alpha_S(\mu)} =
   1 - \beta_0 {\alpha_S(M_Z)\over 2 \pi} \ln \left( {M_Z\over \mu}
\right) ,
\label{as}
\ee
in the leading logarithmic approximation where
$\beta_0 = 11 - {2\over 3}N_f=23/3\ (N_f=5)$.

In calculating the branching ratios of $b\to s \gamma$ and $b\to s g$
in eqs. (\ref{eq:bsA}) and (\ref{eq:bsg}),
one needs to know the values of the scale $\mu$
and the ratio $m_c/m_b$, which are not well determined.
However, since in our case we only want to find out
the maximum value of $Br(b\to s g)$ that is still allowed,
and since the ratio $R$ does not depend on $m_c/m_b$,
we keep $m_c/m_b$ at some fixed value $\sim 1/3$.
Similarly, we neglect the uncertainties
arising from the CKM mixing elements
in our calculations of the $b\to s$ decay branching ratios.
The main uncertainty is therefore in the scale $\mu$.

We study $b\to sg$ and $b\to s\gamma$ numerically
for different sets of parameters $\xi$ and $M_H$.
We find that a smaller value of $\mu$ gives largest $Br(b\to s g)$.
This is demonstrated in Fig. 1 with $\xi=1\,,\ 2$ and $M_H=m_t=170\ GeV$.
In Figs. 2(a) and (b), we present the branching ratio of
$b\to s g$ decay for $m_t = 170$ GeV and $\mu = 2.5$ GeV
for Models I and II, respectively.
The hatched region to the right is ruled out
by the CLEO upper bound on $Br(\bsg)$ of eq. (1).
We notice that the CLEO limit has excluded most of the parameter space in
the $\xi -M_H$ plane for Model II.
The lower bound of
$Br(b\to s\gamma) > 1.0\times 10^{-4}$ excludes the second hatched region
to the left in Fig. 2(a) for Model I.
We further overlay (shaded) the combined constraints
on CKM mixing matrix (since $m_t$ is fixed here at 170 GeV)
from $\epsilon$ parameter in $K\to\pi\pi$ decay, $B$-$\bar B$ mixing,
and the ratios $|V_{cb}/V_{us}|$ and $|V_{ub}/V_{cb}|$ \cite{Geng}.
This constraint is rather stringent for heavy top,
and is the same for both Model I and II since the top coupling
is common in both models.
The solid, dot and dash  curves in Fig. 2(a)
represent $Br(\bsgl)$ being 0.1\% ($< 0.1\%$ between the two solid lines),
1\% and 8\%, respectively in Model I.
For sake of illustration, however, for Model II
the corresponding lines in Fig. 2(b) are for
$Br(\bsgl)=0.7\%\,,\ 1.0\%$ and 1.5\%, respectively.
  From Fig. 2 we see
that, if the top is heavy as suggested by recent
observation of CDF \cite{CDF},
$Br(b\to sg)$ can at most be of order $1\%$
for both Model I and II.
For Model I, in fact, it would be rather difficult to
go much beyond $0.1\%$.

In Fig. 2(a), we find that
the branching ratio of $\bsgl$ goes to zero inside the two solid lines
due to the destructive interference between the charged Higgs and
the standard model contributions.
$Br(\bsgl)$ in the region to the left of the solid line
increases and reaches the standard model value ($\sim 4.5\times 10^{-3}$)
when $\xi=0$.
Note that in the $\xi - M_H$ plane,
for Model I, the interesting region allowed by the CLEO data
is more or less ruled out by the CKM constraints, while
for Model II, the  variation of $Br(b\to s \gamma)$ is much faster
than that of $Br(\bsgl)$.
Thus, in either case, new improvements on the experimental measurement
of the $\bsg$ decay will imply no significant changes on the limits
of $Br(\bsgl)$.


An intriguing possibility still exists for
$M_H +m_b < m_t < M_W + m_b$ \cite{Hou-top},
allowed by all known constraints, including $b\to s\gamma$
and $B$-$\bar B$ mixing.
When top is light, substantial charged Higgs contributions may
in fact be called for. If one takes the heavy quark production signal
observed by CDF \cite{CDF} seriously,
it may actually be the fourth generation $t^\prime$ quark.
In that case, all loop
effects are subject to GIM cancellation, and can be made ineffective.
The crucial point, however, is that
$t\to bH^+$ overwhelms $t\to bW^*$ in this domain,
and can allow the top quark to elude past searches
at hadronic colliders.
For Model I, assuming that the heavy quark
seemingly observed by CDF \cite{CDF}
does not dominate in the loop processes,
we find that this can happen for $\xi=1/\tan\beta \sim 2$.
We illustrate in Fig. 3(a) the allowed region for $Br(b\to sg)$
for $M_H < m_t = 70$ GeV.
Although there is still no large enhancement,
the CKM constraint is now more forgiving,
and a $3\%$ branching ratio for $b\to sg$ is possible.
This cannot be considered small when compared with the
standard model expectation of order $10^{-3}$ \cite{Ciuchini}.
For Model II, $b\to s\gamma$ provides a very stringent constraint,
and in particular it is difficult to evade
the direct search for $t\to bH^+\to b\tau^+\nu$ \cite{tbH}.
Nevertheless, combining the two constraints, it is found \cite{tbH2}
that the region $\xi=1/\tan\beta \sim 1$ is allowed for having a light top
decaying
via charged Higgs (which does not decay dominantly via $\tau\nu$),
especially when one takes into account all the possible sources of
errors in making estimates.
We plot in Fig. 3(b), with same notation as in Fig. 2(b),
the expected $Br(b\to sg)$ that may still be allowed for
Model II. The solid line corresponding to $Br(b\to sg) = 0.7\%$
now falls outside of the figure.
We find that the maximum value of $Br(b\to sg)$
is about 0.9\%, which is indeed smaller than the case for Model I.

As stressed in ref. \cite{GH}, $b\to sg$ at the  $10\%$ level or higher
could account for the apparent discrepancy on $Br(b\to e\nu + X)$
between experiment and theory.
We find that this possibility is quite definitely ruled out,
by the combined limits of $b\to s\gamma$ and CKM matrix, 
especially if one takes the CDF heavy quark production signal
as due to the top quark.
However, in case the top is actually light (and CDF signal is
either faked or due to {\it new} heavy quarks),
$b\to sg$ could still be at $3.5\%$ level.
Since the other possibility for suppressing $Br(b\to e\nu + X)$
by having $b\to \tau\nu + X$ at $10\%$ level or higher
is also  ruled out by ALEPH collaboration \cite{ALEPH},
the two Higgs doublet models cannot help alleviate
the inclusive semileptonic $b$ decay problem.
Perhaps one would have to opt for large $\alpha_S(M_Z)$ (of order 0.13)
and a low $\mu$ scale ({\it e.g.} $\mu \sim m_b/2$) for $B$ decay processes,
as suggested by Altarelli and Petrarca \cite{AP}.
This would imply that $\alpha_S > 0.3$ for $B$ decay processes.

%
%
\acknowledgments

This work is supported in part
by the National Science Council of
the Republic of China
under grants NSC-83-0208-M-007-118 (C.Q.G),
NSC-83-0208-M-007-117 (P.T), and
NSC-83-0208-M-002-023 (W.S.H).
W.S.H. wishes to thank the CERN Theory Division for
hospitality and a stimulating environment.


\newpage
\def\pl#1#2#3{
     {\it Phys.~Lett.~}{\bf #1B} (19#2) #3}
\def\zp#1#2#3{
     {\it Zeit.~Phys.~}{\bf #1} (19#2) #3}
\def\prl#1#2#3{
     {\it Phys.~Rev.~Lett.~}{\bf #1} (19#2) #3}
\def\pr#1#2#3{
     {\it Phys.~Rev.~ }{\bf D#1} (19#2) #3}
\def\np#1#2#3{
     {\it Nucl.~Phys.~}{\bf B#1} (19#2) #3}
\def\ib#1#2#3{
     {\it ibid.~}{\bf #1} (19#2) #3}

\relax
\newpage

\vskip -1cm
\figure{The ratio R as a function of the scale $\mu$ in
        (a) Model I and (b) Model II.
        The solid and dash curves are for $\xi=2$ and 1, respectively.}

\figure{Branching ratio of
        $b\to s g$ decay for $m_t = 170$ GeV and $\mu = 2.5$ GeV
        for (a) Model I and (b) Model II.
        The hatched region to the right is ruled out
        by the CLEO upper bound on $Br(\bsg)$,
        while a second region to the left
        in (a) is excluded by imposing a lower bound of
        $Br(b\to s\gamma) > 1.0\times 10^{-4}$.
        The shaded region is forbidden by constraints on
        the CKM matrix.
        The solid, dot and dash  curves
        represent $Br(\bsgl)$ being 0.1\% ($< 0.1\%$ between
        the two solid lines), 1\% and 8\% in (a),
        while in (b) they correspond to
        $Br(\bsgl)=0.95\%\,,\ 1.0\%$ and 1.3\%, respectively.}

\figure{Same as Fig. 2 but with $m_t=70\ GeV$.}


\begin{references}
\addcontentsline{toc}{chapter}{Bibliography}

\bibitem{CLEO}
CLEO Collaboration, R.~Ammar et. al.,
{ Phys. Rev. Lett.} {\bf 71} (1993) 674.

\bibitem{Hewett-r} For a recent review, see
J.L. Hewett, {\it SLAC-PUB-6521}, May 1994.

\bibitem{CLEO2}
E. Thorndike (CLEO Collaboration),
talk presented at XXVII International Conference on
High Energy Conference (ICHEP),
July 21 -- 27, 1994, Glasgow, Scotland.

\bibitem{qcdi} S.~Bertolini, F.~Borzumati and A.~Masiero,
{ Phys. Rev. Lett.} {\bf 59} (1987) 180;\\
               N.G.~Deshpande {\it et al.}, 
{ Phys. Rev. Lett.} {\bf 59} (1987) 183.

\bibitem{Buras}
A.J. Buras, $et\ al.,$ {\it MPI-Ph/93-77};    
and references therein.

\bibitem{Ciuchini}
M. Ciuchini. $et\ al.,$ {\it CERN-TH 7283/94};
and talk presented by M. Ciuchini at XXVII ICHEP; 
and references therein.
%
%
%
%

%
\bibitem{HW}
W.S. Hou and R.S. Willey, { Phys. Lett.} {\bf B20
2} (1988) 591;
{ Nucl. Phys.} {\bf B326} (1989) 54.

\bibitem{GH}
B. Grzadkowski and W.S.  Hou, { Phys. Lett.} {\bf B27
2} (199
1) 383. 

\bibitem{benx-expt} K. Hikasa {\it et al.}, the Particle Data Group,
{ Phys. Rev}. {\bf D45} (1992) S1.

\bibitem{benx-th}
See, $e.g.$, I. I. Bigi {\it et al.},
{ Phys. Lett.} {\bf B323} (1994) 408;
and references therein.

\bibitem{GW}
S.L. Glashow and S. Weinberg, { Phys. Rev.} {\bf D15} (1977) 1958.

\bibitem{hunter}
Cf. J.F. Gunion $et\ al.,$ {\sl The Higgs Hunter's Guide},
Addition-Wesley Pub., 1990.

\bibitem{c24} N. Cabibbo, G. Corb\`o and L. Maiani,
{ Nucl. Phys.}  {\bf B155} (1979) 93.

\bibitem{c25} M. Jezabek and J.M. K\"uhn,
{ Nucl. Phys.}  {\bf B320} (1989)
20 and  {\bf B314} (1989) 1.

\bibitem{Ciu} M.~Ciuchini {\it et al.},
                        { Phys. Lett.} {\bf B316} (1993) 127.



\bibitem{wise}
B. Grinstein and M.B. Wise, { Phys. Lett.} {\bf B201} (1988) 274.


\bibitem{Ali-mu}
A.~Ali, C.~Greub and T.~Mannel, ``Rare $B$ decays in the
                Standard Model'', in Proc. ECFA Workshop on B-Meson
                Factory, eds. R.~Aleksan and A.~Ali, DESY, March 1993.

\bibitem{Geng} Cf. G. B\'elanger, C.Q. Geng and P. Turcotte,
{ Phys. Rev.}
{\bf D46} (1992) 2950.

\bibitem{CDF} CDF Collaboration, F. Abe et al.,
{\it FERMILAB-PUB-94/097-E}.

\bibitem{Hou-top}
W.S. Hou, { Phys. Rev. Lett.} {\bf 72} (1994) 3945.
%
\bibitem{tbH}
CDF Collaboration, F. Abe {\it et al.}, { Phys. Rev. Lett.}
{\bf 72} (1994) 1977.
%
\bibitem{tbH2} W.S. Hou {\it et al.},
{\it NTUTH-94-15} and {\it NHCU-HEP-94-16}, July 1994.

\bibitem{ALEPH} ALEPH Collaboration, D. Buskulic {\it et al.},
{ Phys. Lett.} {\bf B298} (1993) 479.

\bibitem{AP} G. Altarelli and S. Petrarca,
{Phys.Lett.} {\bf B261} (1991) 303.


\end{references}
\end{document}